\documentclass[prl,a4paper]{revtex4}

\usepackage{amsmath,amssymb}
\usepackage{epsfig}

\newcommand{\numero}[1]{\noindent #1.~}
\newcommand{\la}{\lambda}
\newcommand{\be}{\beta}
\newcommand{\pa}{\partial}
\newcommand{\f}{\frac}
\newcommand{\beg}{\begin{equation}}
\newcommand{\bi}{\begin{itemize}}
\newcommand{\ee}{\end{equation}}
\newcommand{\ei}{\end{itemize}}
\newcommand{\ii}{\item}
\newcommand{\g}{\gamma}

\newcommand{\Ga}{\Gamma}
\newcommand{\de}{\delta}

\def\ab{\f {\alpha_s N_c}{\pi}}

\def\kb{\bar k^2}
\def\ka{\kappa}

\begin{document}
\title{Spin-glass model of QCD near saturation}

\author{Robi Peschanski}
\email{pesch@spht.saclay.cea.fr}
\affiliation{Service de physique th{\'e}orique, CEA/Saclay,
  91191 Gif-sur-Yvette cedex, France\footnote{%
URA 2306, unit\'e de recherche associ\'ee au CNRS.}}

\begin{abstract}
We establish a connection between the  cascading of gluon momenta modeled within 
the diffusive approximation of the Balitsky-Fadin-Kuraev-Lipatov kernel and the 
thermodynamics of directed polymers on a tree with disorder. Using known results 
on the low-temperature spin-glass phase of this statistical mechanic problem we 
describe the dynamical phase space of gluon transverse momenta near saturation 
including its fluctuation pattern. It exhibits a  nontrivial clustering 
structure, 
analoguous to ``hot spots'', whose distributions are derived and possess 
universal 
features in common with other spin-glass systems.
\end{abstract}

\maketitle

\vspace{.1cm}  {\bf \numero{1}}{\it Saturation} in QCD is expected to occur when 
parton densities inside an hadronic target are so high that their wave-functions 
overlap. This is expected from the high-energy evolution of deep-inelastic 
scattering governed by the Balitsky~Fadin~Kuraev~Lipatov (BFKL) kernel 
\cite{bfkl}. The BFKL evolution equation is such that the number of gluons of 
fixed size increases exponentially and would lead, if not modified, to an 
average 
violation of unitarity. Saturation  \cite{GLR,CGC} is characterized by a typical 
transverse momentum scale $Q_s(Y)$, depending on the overal rapidity of the 
reaction, at which the unitarity bound is reached by the BFKL evolution of the 
amplitude. The problem we want to address here is the 
characterization of the gluon-momentum distribution {\it near saturation}. This 
study 
may help analyzing  the Color Glass Condensate (CGC) which has been proposed 
\cite{McLerran:2004fg} as a description of the QCD  phase at full saturation. As 
we 
shall see, our aim in the present paper is to fully understand the spectrum of 
transverse momentum fluctuations among the gluons which are generated by the 
BFKL 
evolution in rapidity, {\it near saturation}.

Indeed, one of the major recent challenges in QCD saturation is the problem of  
taking into account the r\^ole of fluctuations which are present in the BFKL 
evolution, i.e. in the {\it dilute} regime. Quite surprisingly, the 
fluctuations in this region modify the overall approach to saturation by a 
mechanism which has been understood as the formation of traveling-wave  
solutions 
of nonlinear differential equations. If at 
first one neglects the fluctuations (in the  mean-field approximation), the 
effect of saturation on a dipole-target amplitude is described by the nonlinear 
Balitsky-Kovchegov \cite{Balitsky} (BK) equation, where a nonlinear damping term 
 adds to the BFKL equation. As shown in \cite{munier}, this 
equation falls into the universality class of the Fisher and 
Kolmogorov~Petrovsky~Piscounov (F-KPP) nonlinear equation \cite{KPP} which
admits asymptotic traveling-wave solutions of ``pulled-front'' type \cite{wave}. 
In the QCD case, this means that they are driven by the gluons of higher 
momentum. 
Hence the overall behaviour of the solution is driven by the high-momentum tail, 
which is precisely in the ``dilute'' domain of the BFKL regime, far from the 
range which is dominated by saturation. The exponential behaviour of the BFKL  
evolution quickly enhances the effects of the tail towards a region where 
finally the nonlinear damping regulates both the traveling-wave propagation and 
structure.

In these conditions it was realized, for ``pulled-front'' traveling waves 
\cite{brunet} and in QCD \cite{Iancu:2004es}, that the fluctuations inherent of 
the 
dilute regime may have a surprisingly large effect on the overall solution of 
the nonlinear equations of saturation. Indeed, a fluctuation in the dilute 
regime 
may grow exponentially and thus modify its contribution to the overall 
amplitude. 
Hence, in order to enlarge our understanding of  the QCD evolution with 
rapidity, 
it seems important to give a quantitative description of the fluctuation pattern 
generated by the BFKL evolution equations for the set of cascading dipoles (and 
thus gluons) {\it near saturation}, which is the goal of our paper. We shall 
work 
in the leading order in 
$1/N_c,$ where the QCD dipole framework is valid 
\cite{Nikolaev:1991ja,Mueller:1993rr}. 

The equations for the generating 
functionale of moments of the QCD dipole distributions have been derived 
already some time ago \cite{Mueller:1993rr} and they have been used in 
instructive 
numerical simulations of onium-onium scattering \cite{salam} and also applied 
recently in diffractive and non-diffractive  deep-inelastic scattering 
\cite{Hatta:2006hs}. However,  to our knowledge, a thorough investigation of 
their 
solutions for the pattern of dipoles or equivalently for the event-by-event 
properties of gluon transverse momenta has not yet 
been considered. We will provide an answer to this question in the QCD dipole 
approximation (and moreover in the diffusive approximation of the 1-dimensional 
BFKL kernel, see further on), and the results appear quite rich already within 
this 
approximation scheme. Many aspects we find show ``universality'' features and 
thus 
are expected to be valid also  in the dipole framework beyond the approximations
(going  beyond  the dipole approximation is still a conceptual challenge 
\cite{Iancu:2005dx}).

The strategy of the present paper is to   determine the structure of the 
event-by-event gluon transverse-momentum 
distributions. In the present paper, and as a first step towards a more complete 
solution, we will simplify the problem, focusing on modeling the QCD gluon 
cascading using the diffusive approximation of the BFKL kernel and aiming at a 
quantitative understanding of the transition-to-saturation phase within this 
scheme. 

We show that one can build a mapping between the energy 
evolution of the 
set of gluon momenta (in the 1-dimensional approximation) and the time evolution 
of directed polymers on a tree with disorder. This is obtained through the 
matching of the partition function of the polymer problem with a related 
event-by-event physical quantity defined from the cascading gluon process. Then 
we 
use the 
relation between the cascading evolution around the unitarity limit of the 
amplitude and the  saturation mechanism to find that a spin glass structure of 
the gluon phase space exists at saturation, in direct analogy with the 
low-temperature phase of the directed polymers. We then derive the physical 
consequences of the spin-glass phase for the distribution of gluon transverse 
momenta at saturation in our model of cascading.

The plan of our study is the following. In section 2,  we shall recall the main 
ingredients of the rapidity evolution of the cascading dipoles and 
gluon-momentum 
distributions as driven by the BFKL evolution in the 1-dimensional and diffusive 
approximations. In section 3, we  
introduce the mapping between  the cascading of gluons and the propagation of 
directed polymers on random trees. In  section 4 we recall the nontrivial 
properties of the directed polymer problem, characterized by a spin-glass 
transition at low temperature. In  section 5, we provide a check of the 
consistency of the 
gluon/polymer mapping with the saturation mechanism in the mean field 
approximation. In section 6, we  derive  the gluon phase space properties at 
large rapidity from the spin-glass features. The final section 7 is devoted to 
conclusion and outlook.

\vspace{.4cm}  {\bf \numero{2}}In this section, we present some known results 
on the equations governing the dipole distributions in the BFKL framework 
together 
with those concerning the saturation 
amplitude in the mean-field approximation and relevant for our study. 

\subsection{QCD rapidity evolution}

\begin{figure} [ht]
\begin{center}
\epsfig{file=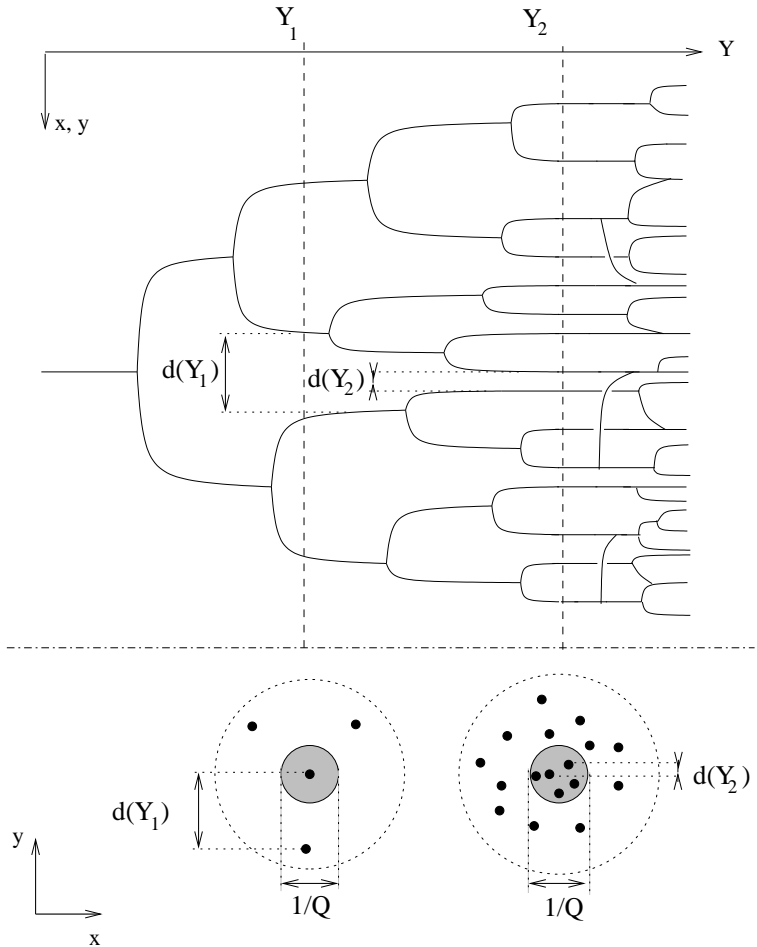,width=10cm}
\end{center}
\caption{{\it BFKL cascading and saturation}.
The  QCD branching process in the BFKL regime and beyond is represented  
along the rapidity axis (upper part). Its 
2-dimensional counterpart in transverse position space is displayed at two 
different rapidities (lower part). The interaction region is represented by a 
shaded disk of size $1/Q$. At rapidity $Y_1$, the interaction still probes 
individual 
dipoles (or gluons), which corresponds to the  exponential  BFKL regime. There 
exists 
a smooth transition to a regime where  the interaction only probes 
groups of dipoles or 
gluons, e.g. at rapidity $Y_2$. This gives a  description of the {\it 
near-saturation} region corresponding to a mean-field approximation, where 
correlations can be neglected. Further in rapidity, $Y> Y_2,$ other dynamical 
effects, such that merging and correlations appear, leading eventually to the 
CGC.}
\label{fig2}
\end{figure}

Let us start by briefly describing the QCD evolution of the dipole distributions 
in 
the 
Balitsky-Fadin-Kuraev-Lipatov BFKL framework 
\cite{Mueller:1993rr,salam,Hatta:2006hs}. 

Formally, let us denote ${\cal Z}_{{\bf v}{\bf w}}[Y,U]$ the generating 
functional 
of moments for the N-uple dipole-probability distributions after a rapidity 
evolution range $Y.$ The dipole probabilities can be deduced from ${\cal 
Z}_{{\bf 
v}{\bf w}}[Y,U]$ by functionally differentiating with respect to the source 
fields 
$U$ and then 
letting $U\to 0.$ the evolution equation satisfied by ${\cal Z}_{{\bf v}{\bf 
w}}[Y,U]$ reads
\beg
\f{\partial{\cal Z}_{{\bf v}{\bf w}}}{\partial Y}\equiv 
\int_z d^2z\ {\cal K}({\bf v},{\bf w};{\bf z})\ 
\left\{{\cal Z}_{{\bf v}{\bf z}}{\cal Z}_{{\bf z}{\bf w}}-{\cal Z}_{{\bf v}{\bf 
w}}\right\}\ ,
\label{equation}
\ee
where
\beg
{\cal K}({\bf v},{\bf w};{\bf z}) = \ab \ \f {\left({\bf v}-{\bf 
w}\right)^2}{\left({\bf v}-{\bf z}\right)^2 \left({\bf z}-{\bf w}\right)^2}\ 
\label{Kernel}
\ee
is the BFKL kernel \cite{bfkl} describing the dissociation vertex of one dipole 
$({\bf v},{\bf w})$ into two dipoles at 
$({\bf v},{\bf z})$ and $({\bf z},{\bf w}),$ where ${\bf v},{\bf w},{\bf z}$ are 
arbitrary 2-dimensional transverse space coordinates .

From Eq.\eqref{equation}, we will mainly retain its physical meaning. The 
functionale  structure of Eq.\eqref{equation} describes a 
2-dimensional tree structure of dipoles in transverse position space evolving 
with 
rapidity. Let us, for instance, focus on the rapidity 
evolution starting from one massive $q\bar q$ pair or onium 
\cite{Mueller:1993rr}, see Fig.\ref{fig2}. 
At each 
branching vertex, the wave function of the onium-projectile is described by a 
collection of color dipoles. The dipoles split with a probability per unit of 
rapidity defined by the BFKL kernel 
\eqref{Kernel}. The BFKL rapidity evolution generates a cascade of dipoles. Each 
vertex of the cascade is governed in coordinate space by the 2-dimensional BFKL 
kernel (\ref{Kernel}), determining both the energy and transverse 
position 
space evolution of the cascade, as sketched in Fig.\ref{fig2}. As we shall see 
later on, this cascading property can be converted in a similar cascade of 
gluons 
in momentum space.

At this stage, it is interesting  to note \cite{Hatta:2006hs} the mathematical 
similarity of \eqref{equation} with the BK equation \cite{Balitsky}, as  
verified by the S-Matrix element of the 
dipole-target 
amplitude {\cal S}({\bf v},{\bf w},Y) in the mean-field approximation. The 
equation 
is indeed formally the same if one replaces term by term the functional ${\cal 
Z}_{{\bf v}{\bf w}}[Y,U]$ by the function ${\cal S}({\bf v},{\bf w},Y)$  
representing the S-Matrix element.

\subsection{Gluons in the 1-dimensional and diffusive approximation}

As an approximation of  the 2-dimensional formulation of the BFKL kernel 
\eqref{Kernel}, obtained when 
one neglects the impact-parameter dependence, we shall 
restrict our analysis in the present paper  to the 1-dimensional reduction of 
the problem to the transverse-momenta squared $k_i^2$ of the cascading gluons.  
After Fourier transforming to 
transverse-momentum space, the leading-order BFKL kernel \cite{bfkl} defining 
the rapidity evolution in the 1-dimensional approximation is known 
\cite{Balitsky} to act in transverse momentum space as a  
differential operator of infinite order 
\beg
\chi(-\partial_l) \equiv  2\psi(1)-\psi(-\partial_l)-\psi(1+\partial_l)
\label{eq:lkernel}
\end{equation}
where $l=\log k^2$ and $Y$ is the rapidity in units of the fixed  coupling 
constant $\ab$.

In the sequel, we shall restrict further our analysis to the diffusive 
approximation of 
the BFKL kernel. We thus expand the BFKL kernel  to second order around some 
value $\gamma_c$
\begin{equation}
\chi(\gamma) \sim \chi_c + 
\chi'_c(\gamma-\gamma_c)+{\scriptstyle \frac{1}{2}}\chi''_c(\gamma-\gamma_c)^2
             = A_0 - A_1\gamma + A_2\gamma^2\ ,
\label{eq:coefsai}
\end{equation}
where $\gamma_c$ will be defined in such a way to be relevant for the {\it 
near-saturation} region of the BFKL regime.

Within this diffusive approximation, it is easy to realize that the first term 
($A_0$) is responsible for the exponential increase of the BFKL regime while 
the third term ($A_2$) is a typical diffusion term. The second term ($A_1$) is 
a ``shift'' term since it amounts to  a rapidity-dependent redefinition of the 
kinematic 
variables, as we shall see.
 
In Eq.\eqref{eq:coefsai},
$\gamma_c$ is chosen in order to ensure the validity of the kernel 
\eqref{eq:coefsai} in the transition region from the BFKL regime towards 
saturation.
Indeed, the derivation of asymptotic solutions of the BK equation 
\cite{munier} leads to consider  the 
condition 
\begin{equation}
\chi(\gamma_c) = \gamma_c\ \chi'(\gamma_c)\ 
\label{eq:condition}
\end{equation}
whose solution determines $\gamma_c.$

This condition applied to the kernel formula \eqref{eq:lkernel} gives $\gamma_c 
=\sqrt {A_0/ A_2}\approx 0.6275...$ and  
$\{A_0,A_1,A_2\}\approx 
\{9.55,25.56,24.26\}.$
It is worth noting that a more 
general choice of the $A_i$'s can be considered for the forthcoming discussion, 
leaving 
the possibility of considering also ``effective'' kernels 
\cite{Peschanski:2005ic}, eventually taking into account next leading 
corrections
and thus other values of $\gamma_c.$ 

\subsection{Saturation in the mean-field approximation}

As already noticed \cite{Hatta:2006hs}, the 2-dimensional BK equation describing 
saturation in the 
mean-field approximation is formally similar to \eqref{equation} by substituting 
${\cal Z}$ by ${\cal S},$ the S-matrix element of the scattering amplitude. In 
the 
1-dimensional approximation which we will use, the BK equation reads
\beg
\partial_Y T(l,Y)  = \chi(-\partial_l)\ T(l,Y) -  T^2(l,Y)\ , 
\label{eq:BK}
\end{equation}
with the same notations as in \eqref{eq:lkernel}.

In the 1-dimensional approximation, the Fourier transform of the dipole-target 
scattering  amplitude $T(l,Y)$ is related to the unintegrated gluon distribution 
in 
the target (by a simple convolution). Applying only the linear operator 
\eqref{eq:lkernel} to 
$T(l,Y)$  leads to an exponential increase of the gluon density with rapidity 
characteristic of the BFKL regime. The damping term in Eq.\eqref{eq:BK} is 
responsible for the saturation mechanism in the mean-field approximation.

There exists an already long literature on the properties of the BK equation 
(see, 
for instance, \cite{CGC,Balitsky,munier}). For further purpose, we remark that  
many results which can be obtained  from the solutions of the BK 
equation can already be derived from the BFKL evolution when only the linear 
operator in Eq.\eqref{eq:BK} is retained. The idea 
\cite{Iancu:2002tr,Mueller:2002zm} is to 
consider the extension of the BFKL evolution in the rapidity region when the 
unitarity limit is reached and assume a smooth transition to saturation. This 
amounts to study the BFKL regime alongside the saturation line defined by the 
saturation scale $Q_s(Y)$ in the 2-dimensional $k^2,Y$ plane.

Indeed, one 
obtains a significant part \cite{Iancu:2002tr,Mueller:2002zm} of the 
results obtained 
\cite{munier} by a direct derivation of the asymptotic solutions of 
the BK equation. 
Let us list some of the most relevant results of this approach based on the 
unitarity bound:
The gluon momentum distribution can be derived at high rapidity and large enough 
$k.$ One finds
\begin{equation}
{T}(l,Y) \propto
\log\left[\frac{k^2}{Q_s^2(Y)}\right]
\left(
\frac{k^2}{Q_s^2(Y)}\right)^{-\sqrt {A_0/ A_2}}
\exp\left\{-\frac{1}{A_2\ Y}
\log^2\left[\frac{k^2}{Q_s^2(Y)}\right]\right\}\ .
\label{nfixedalpha}
\end{equation}
The  {\it saturation scale} ${Q_s^2(Y)}$ is 
given by
\begin{equation}
\log Q_s^2(Y) \propto\ [2\sqrt {A_0 A_2}-A_1]\ Y
-\frac{3}{2\sqrt {A_0/ A_2}}\ \log Y + {\cal O}(1)\ .
\label{qsfixedalpha}
\end{equation}
Furthermore, at large rapidity, the third term of (\ref{nfixedalpha}) and the 
subasymptotic correction in \eqref{qsfixedalpha} become negligible, leaving a 
gluon momentum distribution exhibiting geometric scaling \cite{Stasto:2000er}
\begin{equation} 
{T}(l,Y) \to {T}\left(\tau= \left[\frac{k^2}{Q_s^2(Y)}\right]\right)\ \propto\ 
\tau^{-\sqrt 
{A_0/ 
A_2}}\ 
\log 
\tau \ ; \quad \quad \log Q_s^2(Y) \propto\ [2\sqrt {A_0 A_2}-A_1]\ Y\ .
\label{geometrical}
\end{equation}
The asymptotic results \eqref{geometrical}  can 
be directly obtained from the unitarity bound applied to the solution of BFKL 
equation 
\cite{Iancu:2002tr}, while the subasymptotic results can also be obtained from 
the BFKL equation, but with 
more care taken about 
the boundary conditions \cite{Mueller:2002zm}. 

\vspace{.4cm}  {\bf \numero{3}}Now we introduce the  mapping from the BFKL 
regime of QCD to the thermodynamics of directed polymers. As we have mentionned 
earlier, the BFKL 
rapidity evolution generates a cascade of gluons \cite{Mueller:1993rr}. Each 
vertex of the cascade is governed in coordinate space by the 2-dimensional BFKL 
kernel, determining together the energy and the transverse space evolution of 
the 
cascade, see Fig.\ref{fig2}. In transverse-momentum space and within the 
1-dimensional diffusive approximation (\ref{eq:coefsai}),  we already noticed 
that the BFKL kernel models boils down to a branching, shift  and diffusion  
operator acting in the gluon transverse-momentum-squared space. Hence the 
cascade of gluons  can be put in correspondence with  a continuous branching,  
velocity-shift and  diffusion  probabilistic process whose probability by unit 
of rapidity is defined by the coefficients $A_i$ of \eqref{eq:coefsai}.
 
Let us first introduce the notion of  gluon-momenta ``histories'' $k_i(y).$ They  
register the evolution of the gluon-momenta starting from the unique initial 
gluon momentum $k(0)$ and terminating with the specific $i$-th momentum $k_i,$ 
after successive branchings. They  define  a random function of the {\it running 
rapidity}  $y,$ with $0 \le y \le 
Y,$  
the final 
rapidity range when the evolution ends up (say, for a given total energy). It is 
obvious that two different histories $k_i(y)$ and $k_j(y)$ are equal before the 
rapidity when they branch away from their  common ancestor, see Fig.\ref{fig2}.

We then introduce random paths $x_i(t)$ using formal {\it space} and  {\it 
time} coordinates $x, t\ ,\ 0\le t \le {\Delta t} $  which we  relate to  
gluon-momenta ``histories'' as follows:
\begin{equation}
 \ y = \ \f t{A_0}\ ;\quad 
\quad \log  {k_i^2(y)} \equiv   - \be\ (x_i(t)-x(0)) + ({A_0}-{A_1}) y
\label{eq:mapping}
\end{equation}
where $(A_0-A_1)y$ is a conveniently chosen and deterministic ``drift term'', 
$x(0)$ is an arbitrarily fixed origin of an unique initial gluon and thus the 
same for all susequent random paths. The random paths $x_i(t)$ are 
generated by a continuous
branching and Brownian diffusion process in space-time (cf. Fig.\ref{fig:fkpp}).

As we shall determine later on, the important parameter $\be,$ which plays the 
r\^ole of an inverse temperature for the Brownian process, will be  fixed to
\begin{equation}
\be = \sqrt {\f{2A_2}{A_0}}\ .
\label{eq:mu}
\end{equation}

In fact, the relation \eqref{eq:mu}  will be required by the condition that the 
stochastic process of  random paths describes the BFKL regime of QCD {\it near 
saturation}. Another choice of $\be$ would eventually describe the same 
branching 
process but in other conditions. Hence the condition \eqref{eq:mu} will be 
crucial to determine the QCD phase at saturation (within the diffusive 
approximation).

Let  now introduce the tree-by-tree random function defined as the partition 
function of the random paths system
\begin{equation}
Z(t) \equiv\sum_{i=1}^{n}e^{-\be x_i(t)} = e^{-\be x_0+A_1y}\times \f{ 
\sum_{i=1}^{n} k_i^2(y)}{e^{A_0y}}\ .
\label{eq:Z1}
\ee
Using  the fact that the  total number of momenta in 
one event at rapidity $y$ is $n\approx \langle\  n \ \rangle=e^{A_0 y}$ at large 
$y,$ \eqref{eq:Z1} takes the form
\begin{equation}
Z =  \lambda\ e^{A_1 
y}\times \kb(y)\ ,
\label{eq:Z2}
\end{equation}
where $\f 1n\sum_{i=1}^{n} k_i^2(y)\equiv \kb(y)$ is the  {\it event-by-event} 
average over gluon momenta at rapidity $y$ and  $\lambda =\ e^{-\be x_0}$ is an 
arbitrary constant. Note that one has to distinguish $\langle\  \cdots \ 
\rangle,$ i.e.  the sample-to-sample average, from   $\overline{\ \cdots\ }$ 
which 
denotes the average made over only {\it one} event. 

$Z(t)$ is an event-by-event random function. The physical properties are 
obtained 
by averaging various observables over the events. Note that the distinction 
between averaging over one event and the  sample-to-sample averaging appears 
naturally   in the statistical physics problem in terms of ``quenched'' 
disorder: 
the time scale associated with the  averaging over one random tree structure is  
different (much shorter) than the one corresponding to the averaging over random
trees. 

In parallel with the statistical physics problem \cite{Derrida}, it is 
convenient 
 to introduce the generating functional $G(t,x)$ of its event-averaged moments 
$\langle\   Z^p \ \rangle,$ namely
\begin{equation}
G(t,x)\equiv \sum_p \f 1{p!}\ \langle\  \left[-e^{\ \be x}\ 
Z(t)\right]^p\rangle\ 
= 
\langle\  \exp\left[-e^{-\be x}\ Z(t)\right]\ \rangle\ = \langle\  
\exp\left[- \lambda\ {\kb(y)}/{\ka^2}\right]\ \rangle\ ,
\label{eq:G}
\end{equation}
where 
\begin{equation}
y=\f t{A_0},\quad \log \ka^2 =\be x +\f{A_1}{A_0}\ t\ . 
\label{eq:dumb}
\end{equation}
The moments $\ \langle\  
Z^p(t)\rangle\ $ are easily recovered by differentiation over $x$ or $\log 
\ka^2$  at fixed $t.$

It is interesting to note the final form of \eqref{eq:G} showing a  
similarity \cite{Hatta:2006hs} with a unitary S-Matrix amplitude if $\ka$ is 
interpreted as the 
scale  of a transverse-momentum probe, such as depicted in Fig.\ref{fig2}. This  
similarity will be useful for the derivation of saturation properties from 
the BFKL regime in the unitarity limit.

\vspace{.1cm}  {\bf \numero{4}}Now that we have established the connection 
\eqref{eq:mapping} 
between 
the 
model of  gluon-momenta submitted to branching, shift and diffusion, and the 
random paths defined by  $x_i(t),$ let us rederive 
the 
properties of QCD gluon cascading {\it near saturation} from the point-of-view 
of 
the
statistical  random-path system. For this 
sake, we recall the corresponding model for directed polymers on a tree, as 
defined in Ref.\cite{Derrida}.

Directed polymers on a random tree are modeled, see Fig.\ref{fig:fkpp}, by the 
time evolution of random 
paths in 1-dimensional space with continuous diffusion and branching (with 
probability per unit of time fixed respectively to $\scriptstyle \f 12$ and 
$1,$ by convention).

Indeed,  contemplating the definitions \eqref{eq:Z1} and \eqref{eq:G} it is 
straightforward to identify $Z$ as the sample-to-sample  {\it partition 
function} 
and $G(t,x)$ as the generating fonctional of its  sample-averaged moments for 
the 
model  \cite{Derrida} of directed polymers on a random tree.
\vspace{.5cm} 
\begin{figure}[ht]
\epsfig{file=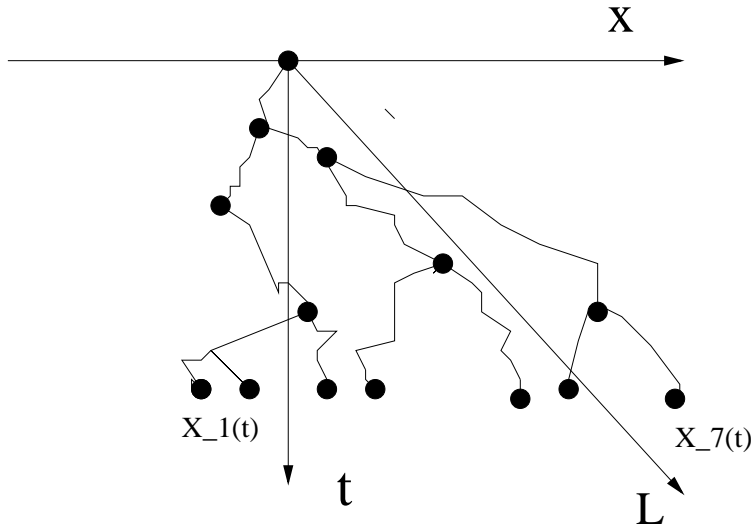,width=10cm}
\caption{{\it Branching diffusion model for polymers}. The coordinates 
$x_1(t)\cdots x_7(t)$ correspond to the random paths along the tree in the 
$x,t$ phase-space. The oblique axis is for $L= \be x +{A_1}/{A_0}\ t,$ which 
takes into account the ``time-drift'' in the mapping to the QCD problem.}
\label{fig:fkpp}
\end{figure}

Let us list some relevant properties of the polymer problem 
\cite{Derrida}.
\bi
\ii
 $u(t,x)\equiv 1-G(t,x)$ verifies the Fisher and 
Kolmogorov~Petrovsky~Piscounov (F-KPP) 
equation \cite{KPP}
\begin{equation}
\f{\pa}{\pa t}u(t,x)={\f 12}\  \f{\pa^2 u(t,x)}{\pa x^2}+u(t,x)-u^2(t,x)\ .
\label{eq:u}
\end{equation}
$u(t,x)$  admits traveling-wave solutions at large time $t$ \cite{wave}
\begin{equation}
u(t,x)\sim u(x-m_{\be}(t)) \approx u(x-c(\be)t)\ ,
\label{eq:u1}
\end{equation}
where  the wave speed $c(\be)$ is governed by the initial condition 
$u(t_0,x) \sim e^{-\be x}$ at large $x.$ The range of values of $c(\be)= 1/\be 
+ \be/2$ is 
bounded from below by $c(\be_c)\equiv c={\sqrt 2},$ where the wave velocity 
remains 
fixed whenever $\be \ge \be_c ={\sqrt 2}.$ $c$ and $\be_c$ are usually called, 
respectively, the critical velocity and the critical slope of the waves. In the 
``supercritical'' regime $\be \ge \be_c ,$ the wave front and its speed remain 
``frozen'' and 
\begin{equation}
m_{\be}(t)= c\ (t - \f 34 \log t) + {\cal O}(1)\ .
\label{eq:m}
\end{equation}
\ii
The  {\it free energy} of the polymer system at equilibrium, defined  by
\begin{equation}
F=-\f 1{\be} \langle\ \log Z(t)\ \rangle\equiv\ -m_{\be}(t)\ ,
\label{eq:F}
\end{equation}
has a  ``frozen'' density $F/t\to -c$ for $\be \ge \be_c= {\sqrt 2}$  
indicating the existence of a 
phase transition at $\be_c.$  This is related to the dynamical transition  
$c(\be)\to c$ in the traveling-wave properties \cite{wave}.
\ii
Introducing the {\it tree-by-tree}  free energy spectrum
\begin{equation}
f(t) = -\f 1{\be}  \log Z(t) + \f 1{\be} \langle\  \log Z(t) \ \rangle \ ,
\label{eq:f}
\end{equation}
its probability distribution 
${\cal P}_{\be}(f)$  around the average free energy $F$ is found by deriving 
asymptotic expressions for the moments \cite{Derrida} to satisfy 
\begin{eqnarray}
{\cal P}_{\be}(f) &\sim& -f\ e^{{\sqrt 2} f}\ \ \ \ \ \ \ \ \  {\rm for}\ 
\ \ 
\  f\to -\infty \nonumber \\
{\cal P}_{\be}(f) &\sim&  e^{-[(2-{\sqrt 2}) f]}\ \ \ \ {\rm for}\ \ \ \  
f\to 
\infty 
\ ,
\label{eq:Proba}
\end{eqnarray}
where we quoted only the distribution obtained for the supercritical regime 
$\be \ge 
\be_c .$ This means that the fluctuations of the free energy around its event 
average remain finite even at asymptotic time, i.e. there is no thermodynamical 
self-averaging, 
demonstrating that the phase transition is {\it not} similar to an ordinary 
liquid-gas  phase transition.
\ii
The corresponding phase transition is  towards a {\it spin-glass} phase 
characterized by a non-trivial structure in ``valleys'' or ground state 
minima, which can be explored quite in detail. We shall describe this structure 
in the 
next section, since it will give, through matching with the gluon cascading
problem, a description of the gluon-momentum phase at saturation within the 
diffusive mean-field scheme.
\ei


\vspace{.1cm}  {\bf \numero{5}}Let us make now the connection of the 
directed-polymer 
properties with the description of the gluon-momentum phase {\it near  
saturation}. 

Our aim is to  study the structure of the gluon-momentum spectrum when the 
system 
reaches saturation. We thus start with  a simple initial state (e.g. an onium or 
one gluon) 
and, the rapidity increasing, the state develops into a cascading structure of 
gluons governed by the model that we assume. It consists in random tree 
structures generated  by branching, shifting and diffusing, as previously 
described. As we have discussed in introduction, the rapid exponential increase 
of the number of gluons generates  reinteractions which damp the 
exponential increase and cause the saturation phenomenon \cite{CGC}.

In the simple mean-field approximation which neglects the dynamical correlations 
\cite{Balitsky}, the saturation mechanism is known to  be studied from the BFKL 
properties in the following approximation \cite{Iancu:2002tr}. When the 
exponential increase of the gluon density reaches the unitarity limit of the 
amplitude, the saturation effects set in. This defines a kinematical region in 
the $k,Y$ plane around the saturation scale $Q_s(Y)$ where the BFKL regime is 
expected to smoothly connect the saturation region. Indeed, this expectation has 
been verified for the amplitude, where many aspects of the saturation features 
rigorously found \cite{munier} for the solutions of  the mean field BK 
saturation 
equation have been already derived from the BFKL regime considered near the 
unitarity limit \cite{Iancu:2002tr}.

We shall now try and extend this approach of saturation, which provided already 
valuable results on averaged observables, to the event-by-event spectrum 
of gluon transverse momenta. In other words, we study momenta of the order of  
the saturation scale, i.e. $k_i^2 = {\cal O}(Q^2_s).$ This is nothing else than 
the neighbourhood of the kinematic region where ``universal'' traveling-wave 
symptotic solutions of the BK equation exist \cite{munier}. We thus expect from 
the studies of Ref.\cite{munier} that the properties we shall derive will be 
independent of initial conditions (provided they are leading to a 
``supercritical'' 
regime), of the precise form of the kernel (beyond the diffusive approximation) 
and even of the precise form of the non-linearities \cite{wave}. 

A comment of caution is nevertheless in order here, concerning the  domain of 
validity of our 
appoach. The properties in the region of smaller momenta 
$k_i/ 
Q_s  \ll {1}$ 
will probably depend more on the specific structure of the BFKL kernel beyond 
the 
diffusive approximation while the region with $k_i/ Q_s  \gg {1}$ will keep the 
memory of the initial conditions. They should require a different treatment, 
which is not included in our present work and deserves later study.

Inserting the QCD definitions  (\ref{eq:Z2}) in (\ref{eq:F}) for observables 
after a final rapidity evolution-range $Y$ (or ${\Delta t}$ in time) gives
\begin{eqnarray}
\langle\  \log \kb \ \rangle &\equiv& \langle\  \log Z \ \rangle-\log 
\lambda-A_1 Y= \be \ m_{\be}({\Delta t}) - A_1 Y-\log \lambda \nonumber\\
&=& 2\sqrt{\f 
{A_2}{A_0}}\ 
\left(A_0 Y-\f 34 \log Y\right)- A_1 Y +  {\cal O}(1)\simeq  \log Q_s^2\ .
\label{eq:saturation}
\end{eqnarray}
The second line of Eqs.\eqref{eq:saturation} is obtained by identifying the 
inverse temperature $\be$ with its value defined by the relation \eqref{eq:mu}. 
This the condition to verify $\kb \approx Q_s^2,$ the  identification 
coming from formula \eqref{qsfixedalpha}. This condition is crucial to check the 
main saturation property that the sample averaging gluon momentum stays at the 
value of the saturation 
scale. Hence,
the relations \eqref{eq:saturation}, important for the QCD saturation problem, 
determines the equivalent temperature of the polymer system. Starting from 
\eqref{eq:coefsai} 
and \eqref{eq:mu},  the inverse temperature obtains $\be = 1/\mu \approx 2.25 > 
\be_c={\sqrt 2}$   which is  larger than  the critical value.  As a consequence, 
we 
are 
led  to consider and study the low-temperature phase of the equivalent 
random-path system in order to get information on the QCD phase of gluons {\it 
near saturation}. 

As a further check of  the validity of the random 
cascading model in the ``unitarity limit'' region, where the average momentum is 
consistently found to be determined by the saturation scale, let us discuss the 
distribution around the average.

The event-by-event mean momentum $\kb$ at rapidity $Y$ can be read off  from 
Eq.\eqref{eq:f} 
as being
\begin{equation}
\log \left[\f {\kb}{Q_s(Y) ^2}\right] \equiv -\be\ f({\Delta T}) = - \sqrt{\f 
{2A_2}{A_0}}\ f(A_0 Y)
\label{eq:saturation1}
\end{equation}
and, from the first line of \eqref{eq:Proba}, when $e^{-\be f} \sim  
{\kb}/{Q_s^2(Y)}$ is large, its  probability  distribution is given by

\begin{equation}
{\cal P}\left(\log \left[\f {\kb}{Q_s ^2}\right]\right) \ \propto \ 
\log\left[ \f {\kb}{Q_s ^2}\right]\ 
\exp\left\{-\sqrt{\f {A_0}{A_2}} \log \left[\f {\kb}{Q_s ^2}\right]\right\}\ 
.
\label{eq:Proba2}
\end{equation}
By comparison with formula \eqref{geometrical}, the interpretation of Eq.
(\ref{eq:Proba2}) is clear. The probability law (\ref{eq:Proba2}) for the 
event-by-event partition function coincides with the gluon distribution in the 
saturation region and follows geometric scaling. This gives the confirmation, 
through the properties of the directed polymers, that the gluon-cascading model 
near the unitarity limit gives the correct saturation scale and gluon 
distribution. In mathematical terms,  this comes from the fact that the equation 
for the generating function of moments for the model \eqref{eq:u} is in the same  
(FKPP) universality class than the  BK equation \cite{munier}.

Both properties \eqref{eq:saturation},\eqref{eq:Proba2} prove the consistency of 
the model with the properties expected from saturation. We shall then look for 
other properties of the cascading gluon model. It is important to realize that 
this consistency 
fails for any other choice of the parameter $\be$ different from \eqref{eq:mu}. 
This justifies 
a-posteriori the identification of the equivalent temperature of the system in 
the 
gluon/polymer mapping framework.

\vspace{.2cm}  {\bf \numero{6}}Let us now come to the main topics concerning the 
determination of structure of the gluon-momentum phase at saturation in the 
diffusive model.

The striking property of the directed polymer problem on a random tree is the 
spin-glass structure of the low temperature phase. As we shall see this will 
translate directly into a specific  {\it clustering} structure  of gluon 
transverse momenta in their phase near the ``unitarity limit''. Let us divide 
the 
discussion in a few subsections.

i) {\it The equivalent  temperature}. Following the   
relation \eqref{eq:mu} , one is led to 
consider the polymer system at temperature 
$T$ with
\beg
\f {T_c-T}{T_c} \equiv 1-\f {\be_c}{\be} = 1- \g_c\ ,
\label{eq:T}
\ee
where $\g_c$ is the critical exponent. We are thus naturally led to consider the 
low-temperature phase ($T<T_c$), at some distance $T_c-T$ from the critical 
temperature 
$T_c=\scriptstyle{1/{\sqrt 2}}.$ In the language of traveling waves \cite{wave}, 
this corresponds to a ``pulled-front'' condition with ``frozen''and  
``universal'' velocity  and front profile.

ii) {\it The asymptotic phase landscape.} As derived in \cite{Derrida}, the 
phase space of the polymer problem is structured in ``valleys'' which are in the 
same universality class as those of the Random Energy Model (REM) \cite{REM} and 
of the infinite range Sherrington-Kirkpatrick (SK) model \cite{SK}. 

Translating these results in terms of gluon momenta (in modulus), the phase 
space 
landscape consists in event-by-event distribution of clusters of momenta around 
some values $k_{si}^2 \equiv  1/(n_i)\ \sum_{i\in si} k_{i}^2,$ where $n_i$ is 
the cluster multiplicity. The probability weights to find a cluster $s_i$ after 
the whole evolution range $Y$ is 
defined by $$
W_{si}= \f {\sum_{i\in si} k_{i}^2}{\sum_i k_i^2},$$
where the summation in the numerator is over the momenta of gluons within the 
$s_i^{th}$ cluster (see Fig.\ref{valleys}). The normalized distribution of 
weights 
$W_{si}$ thus allows one to study  the  probability distribution of clusters. 
The clustering tree structure, (called ``ultrametric'' in statistical mechanics) 
is 
the most prominent feature of spin-glass systems
\cite{overlap}.

Note again that, for the QCD problem, this property is proved for momenta in the 
region of the ``unitarity limit'', or more concretely in the momentum region 
around the saturation scale. This means that the cluster average-momentum  is 
also such that $k_{si}^2 = {\cal O}(Q^2_s).$ Hence the clustering structure is 
expected to appear in the range  which belongs to the traveling-wave 
front \cite{munier} or, equivalently, of clustering with finite fluctuations 
around  the saturation scale. 

\begin{figure} [ht]
\begin{center}
\epsfig{file=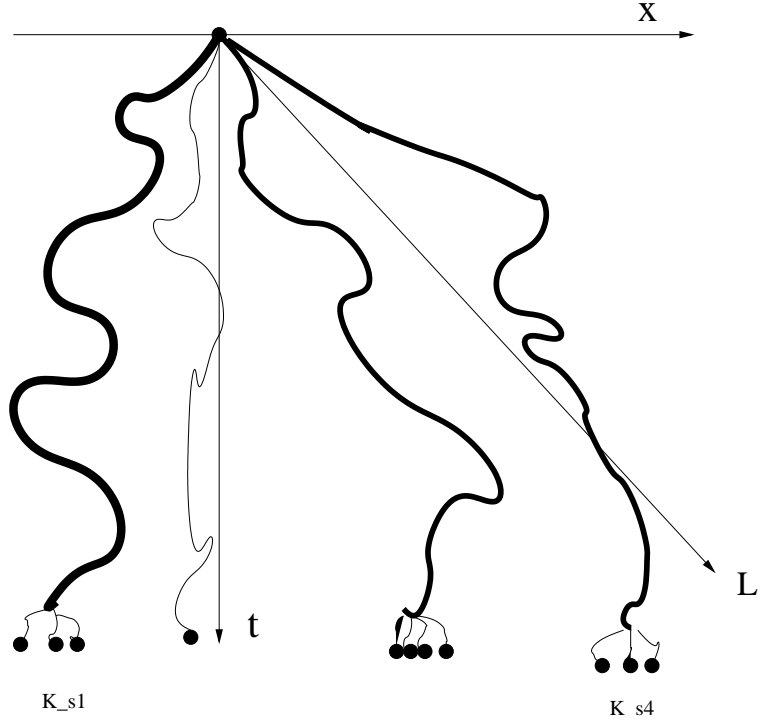,width=10cm}
\end{center}
\caption{{\it The clustering structure}. The drawing represents the $s_1\cdots 
s_4$ clusters near momenta $k_{s1}\cdots k_{s4}$. They branch either near $t\ll 
1$ or 
$(\Delta t-t)\ll 1,$ where $\Delta t$ is the total amount of time evolution.}
\label{valleys}
\end{figure}

iii) {\it The ``overlap'' function.} There exists in statistical physics of 
spin-glasses a well-known function which reveals the 
space-time structure of a spin-glass state in a quantitative way: the 
``overlap'' 
function 
\cite{overlap}. Translating the definitions (cf. \cite{Derrida}) in terms of the 
QCD problem, one  introduces the {\it overlap} between two final gluon momenta 
 $k_i$ and $k_j,$ using their ``histories'' $k_i(y)\ {\rm and}\ 
k_j(y).$ We recall that
$k_i(y)\ {\rm and}\ k_j(y)$ defined as  function of the running rapidity $y$ 
correspond to the two  (not always disjoint) paths histories which finally lead 
to the  
momenta $k_i\ {\rm and}\ k_j$ observed at final (and large) rapidity $Y.$ 

The overlap is defined by the quantity $Q_{ij},$ the fraction  of the total 
rapidity $Y$ for which $k_i(y)=k_j(y), 
\ 
\ \ 0\le Q_{ij}\le 1 \ .$ It is a measure of the rapidity interval (``time'') 
during which they belong to the same branch before splitting.
For a given event, the probability $\tilde {\cal Y}(q)$ of finding an overlap 
$q$ 
can be defined by
\beg
\tilde {\cal Y}(q)\ dq = \f 1{\left\{\sum (k_i^2)\right\}^2}\ \sum_{i,j=1}^n 
k^2_i\ 
k^2_j\ \ 
\Theta(q<Q_{ij}<q\!+\!dq \iff k_i(qY) =k_j(qY)\ {\rm in}\ [q,q+dq])\ ,
\label{eq:P}
\ee
where the characteristic function $\Theta$ restricts the sum to the pairs of 
momenta which where on the same branch in the interval of rapidity $[q,q+dq].$
The result  reads (from Ref.\cite{Derrida})
\beg
\tilde {\cal Y}(q)\ dq\ =\de(q-1)\ {\cal Y}\  + \de (q) \ (1- {\cal Y})\ \equiv 
\de(q-1)\ \sum_{si} W^2_{si}\ + \ \de (q)\ \left(1- \sum_{si} W^2_{si}\right)\ ,
\label{eq:barP}
\ee
where ${\cal Y}=\sum_{si} W^2_{si}$ is an event-by-event indicator of the 
strength of clustering.

The result \eqref{eq:barP} means that, in the thermodynamical limit, the gluon 
momenta branch effectively either near $y=0$ or 
near $y=Y$ for asymptotic $Y$, see Fig.\ref{valleys}. Hence the ``microscopic'' 
branching structure of the model of Fig.\ref{fig:fkpp} at finite $y$, occupies 
only a vanishing (when $Y\to \infty,$ i.e. in the thermodynamical limit)  
fraction of the rapidity evolution, either 
at the 
beginning or at the end but not in the middle nor for a finite fraction of the 
asymptotic rapidity.

The probability distribution of overlaps ${\Pi}({\cal Y})$ is identical to the 
one of the REM and SK models and shares many qualitative similarities with 
other systems possessing a spin-glass phase \cite{toulouse,toulouse1}. Being 
quite
non-trivial analytically, 
it is  more easily defined by its moments
\beg
\langle\  { {{{\cal Y}}}}^{\nu}\ \rangle= \f {(-1)^{n+1}\ T_c/T}{\Ga(2\nu)\ 
\Ga(n\!-\!\nu)}\ \int_0^{\infty} d\rho\ \rho^{n-1-\nu}\ \f{d^n \log 
g(\rho)}{d\rho^n}\ 
\quad\text{for}\quad n\!-\!1<\rho<n\ , 
\label{eq:moments}
\ee
with (see \cite{toulouse,toulouse1})
\beg
g(\rho) = \int_0^{\infty} dv\ v^{-T/T_c-1}\ \left(1-e^{-v-\rho v^2}\right)\ . 
\label{eq:g}
\ee 
The overlap distribution depends only on $T/T_c,$ which confirms its high 
degree 
of universality . One finds, in particular, $\langle\  { {{ {\cal Y}}}}\ \rangle 
= 
1-T/T_c,$ which is the average probability of having  overlap $1,$ while 
$T/T_c$ is 
the probability  of overlap $0$.

The rather involved probability distribution $\Pi({\cal Y})$   is quite 
intringuing, as obtained by inversion of moments 
$\langle\  { {{{\cal Y}}}}^{\nu}\ \rangle$ \cite{toulouse,toulouse1}. It 
possesses apriori an infinite number of singularities at 
${ {{{\cal Y}}}} = 1/n, n\ {\rm integer}.$ It can be seen when the temperature 
is significantly lower from the critical value, see for instance the curve for 
$1-T/T_c = .7$ in Fig.\ref{fig:overlap}. However, the predicted curve for the 
QCD value $1-T/T_c = .3$ is smoother and shows only a final cusp at $W_s=1$ 
within the considered statistics.
It thus seems that configurations with  only one  cluster 
can be  
more prominent than the otherwise smooth generic landscape. However, it is also 
a 
``fuzzy'' landscape since many clusters of various sizes seem to coexist in 
general.
\begin{figure}[t]
\epsfig{file=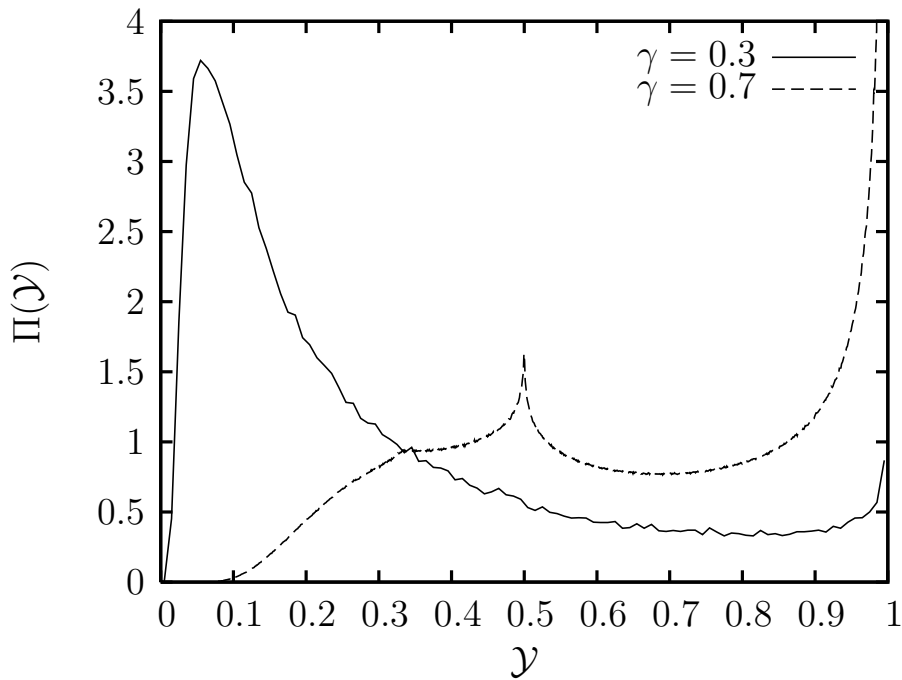,width=12cm}
\caption{{\it The probability distribution of overlaps ${ \Pi}({\cal Y})$.} The 
figure (the simulation is by courtesy from  \cite{private},  using the method of 
Ref.\cite{toulouse1}) is drawn both for  the theoretical QCD
value 
$\g_c = 0.7,$ and for $\g_c = 0.3$ for comparison. The statistics used for the 
simulation is $10^8$ events in $10^3$ bins for $\g_c = 0.3$ and $3.25\ 10^6$ 
events in $10^2$ bins for $\g_c = 0.7.$}
\label{fig:overlap}
\end{figure}

iv) {\it The cluster distribution.} The sample-to-sample distribution 
of ``valleys''  can be explicitely computed by a generalization of the overlap 
function \cite{toulouse,toulouse1} to any power $\epsilon,$ $\sum 
W^{\epsilon}_{si}$, 
which translates into a well-defined distribution of cluster-weight probability. 
For instance one determines:
\begin{eqnarray}
{\Phi}(W) &=& \f {W^{-\g_c-1}\ (1-W)^{\g_c-1}}{\Ga(\g_c)\Ga(1-\g_c)} \nonumber 
\\
{\Phi}(W_1,W_2) &=& \g_c\  \f {(W_1 
W_2)^{-\g_c-1}\ 
(1-W_1-W_2)^{2\g_c-1}}{\Ga(2\g_c)\Ga^2(1-\g_c)} \times \theta(1-W_1-W_2) 
\nonumber \\
{\Phi}(W_1,W_2,\cdots,W_p) &=& \g_c^p\ \Ga(p) \ \f 
{(W_1 
W_2\cdots W_p)^{-\g_c-1}\ (1-\sum_1^p W_i)^{p\g_c-1}}{\Ga(p\g_c)\Ga^p(1-\g_c)} 
\times 
\theta(1-\sum_1^p W_i)\ ,
\label{eq:Pi}
\end{eqnarray}
for, respectively, the average number ${\Phi}(W)$ of  clusters of weight 
$W_{si}\equiv 
W,$ the average number of pairs ${\Phi}(W_1,W_2)$ of clusters in the same 
event with weights $W_{si}\equiv W_1, W_2$, and more generally 
${\Phi}(W_1,W_2,\cdots,W_p)$ the average number of p-uples of clusters with 
weights $W_{si}\equiv W_2, \cdots W_p.$ 

It is interesting to note that smaller is $\g_c=T/T_c,$ smaller is the 
probability to find $W_s \sim 0,$ i.e. no clustering, and higher is the 
probability for having one big cluster $W_s \sim 1.$ As an illustration, the 
phenomenological value $\g_c \approx 0.3$ leads to a more pronounced clustering  
structure than the BFKL value $\g_c \approx 0.6$. The sharp structuration in 
clusters expected for small $\g_c$ can already be seen in the overlap 
distribution of Fig.\ref{fig:overlap}. 

\bigskip
\vspace{.1cm}  {\bf \numero{7}}Let us summarize the main points of our approach. 
We 
investigated the landscape of transverse momenta in gluon cascading around the 
saturation scale at asymptotic rapidity. Limiting our study to a diffusive 
1-dimensional modelization of the BFKL regime of gluon cascading, we make use of 
a mapping on a statistical physics model for directed polymers propagating along 
random tree structures at fixed temperature. We then  focus our study  on the 
region near the unitarity limit where information can be obtained on saturation, 
at least in the mean-field approximation. Our main result is to  find a 
low-temperature spin-glass structure of phase space, characterized by 
event-by-event clustering of gluon transverse momenta (in modulus) in the 
vicinity of the rapidity-dependent saturation scale. The weight  distribution of 
clusters and the probability of  momenta overlap during the rapidity evolution 
are derived.

Interestingly enough the clusters at asymptotic rapidity are branching either 
near the beginning (``overlap $0$'' or $y/Y \ll 1$) or near the end (``overlap 
$1$'' or $1-y/Y \ll 1$) of the cascading event. The probability distribution of 
overlaps is derived and shows a rich singularity structure.

Phenomenological and theoretical questions naturally emerge from this study.

On a phenomenological ground, it is remarkable that saturation density effects 
are not equally spread out on the event-by-event set of gluons; our study 
suggests that there exists random spots of higher density whose distribution may 
possess some universality properties. In fact it is natural to  expect this 
clustering property to be present not only in momentum modulus (as we could  
demonstrate) but also in momentum azimuth-angle. This is reminiscent   of the 
``hot spots'' which were some time  ago \cite{Bartels:1992bx} advocated from the 
production of forward jets in deep-inelastic scattering at high energy (small 
Bjorken variable). The observability of the cluster  distribution through the 
properties of ``hot spots'' is an 
interesting 
possibility.

On the theoretical ground, the questions arise how the pattern of fluctuations 
we 
found influence the onset of saturation and its properties, given the fact that 
fluctuations are of primordial importance in the development of the whole QCD 
process, as we have stressed in the introduction. 

First, the question arises to extend the study to the full BFKL kernel 
\eqref{eq:lkernel}, and in two dimensions, as given initially by 
Eq.\eqref{equation}. Universality properties of the 
solutions of the BK equations in the traveling-wave formalism \cite{munier} and 
in the $2$-dimensional case \cite{marquet} would encourage such a 
generalization. However, some care must be taken of the potential problems which 
can arise from an event-by-event description of the cascading regime. The 
suggestion is \cite{private} to formulate the problem directly in the full 
2-dimensional position space, for which the gluon cascading and the unitarity 
constraints are well defined. 
Some preliminary numerical results are encouraging.

Second, it would be instructive to  incorporate the pattern of 
fluctuations we found in the calculation of the contribution of Pomeron loops to 
the 
scattering amplitude. The  Pomeron loops are found to be important to consider 
and 
are the subject of many studies (see, e.g.
\cite{Iancu:2004es}).  The Pomeron loops are obtained by merging the splitting 
of 
cascading gluons of the projectile with the merging in the target. We think that 
the spin-glass structure of the cascading gluons could lead to useful 
hints 
for dealing with the Pomeron loops.

In conclusion, we think that the already rich structure of the phase space 
landscape of gluon clusters found with the diffusive model of gluon cascading 
opens an interesting way towards a deeper understanding of the QCD phase in the 
saturation regime and could lead to a deeper description the 
Color Glass Condensate.

\vspace{.2cm}
\begin{acknowledgments}

I  would like to  thank Edmond Iancu for inspiring and stimulating remarks and 
Cyrille Marquet and Gregory Soyez for many fruitful suggestions and a careful 
reading of the manuscript. Special thanks are due to Gregory Soyez  for 
providing the simulation of Fig.\ref{fig:overlap}. I am indebted to B.~Derrida 
for his seminal work on 
traveling waves, directed polymers and spin glasses.
\end{acknowledgments}

\end{document}